\documentclass[final]{cvpr}

\usepackage{times}
\usepackage{epsfig}
\usepackage{graphicx}
\usepackage{amsmath}
\usepackage{amssymb}

\usepackage[pagebackref=true,breaklinks=true,colorlinks,bookmarks=false]{hyperref}
\usepackage{multirow}
\newcommand{\xdownarrow}[1]{%
  {\left\downarrow\vbox to #1{}\right.\kern-\nulldelimiterspace}
}

\def\cvprPaperID{11} 
\def\confYear{CVPRW 2021}
\begin{document}

\title{Automotive Radar Interference Mitigation with Unfolded Robust PCA based on Residual Overcomplete Auto-Encoder Blocks}

\author{Nicolae-C\u{a}t\u{a}lin Ristea$^1$, Andrei Anghel$^1$, Radu Tudor Ionescu$^2$, Yonina C. Eldar$^3$\\
$^1$University Politehnica of Bucharest, Romania,\\
$^2$University of Bucharest, Romania,\\
$^3$Weizmann Institute of Science, Israel
}

\maketitle

\begin{abstract}
In autonomous driving, radar systems play an important role in detecting targets such as other vehicles on the road. Radars mounted on different cars can interfere with each other, degrading the detection performance. Deep learning methods for automotive radar interference mitigation can successfully estimate the amplitude of targets, but fail to recover the phase of the respective targets. In this paper, we propose an efficient and effective technique based on unfolded robust Principal Component Analysis (RPCA) that is able to estimate both amplitude and phase in the presence of interference. Our contribution consists in introducing \textbf{r}esidual \textbf{o}ver\textbf{c}omplete \textbf{a}uto-\textbf{e}ncoder (ROC-AE) blocks into the recurrent architecture of unfolded RPCA, which results in a deeper model that significantly outperforms unfolded RPCA as well as other deep learning models. We also show that our approach achieves a faster processing time compared to state-of-the-art fully convolutional networks, thus being a suitable candidate to be deployed on devices embedded on vehicles.
\end{abstract}


\section{Introduction}
\label{sec:intro}

Road safety is a key issue in autonomous driving systems, requiring vehicles to perceive their surroundings. One of the most common proposals made by automotive companies is to employ radar sensors in order to build systems that allow cars to scan the surrounding environment.
Usually, the radar senors used in the automotive industry are frequency modulated continuous wave (FMCW) / chirp sequence radars, which transmit sequences of linear chirp signals. The signals transmitted and received by such sensors provide the means to estimate the distance and the velocity of nearby targets (\eg, vehicles, pedestrians or other obstacles).

The rapidly increasing number of radar sensors \cite{kunert2012eu} unavoidably leads to a higher probability of radio frequency interference (RFI), which generates corrupted and unusable signals. The most common RFI effect involves raising the noise floor by a large margin, to the point where potential targets are completely hidden by noise, thus reducing the sensitivity of target detection methods \cite{brooker2007mutual}. In order to be able to accurately detect targets from signals affected by RFI, the interference has to be mitigated. 
To address this problem, researchers have proposed various techniques ranging from handcrafted approaches \cite{bechter2017dbfmimo, kim2018peer, laghezza2019nxp, mosarim2010d15, uysal2019sync, xu2018orthogonal_noise} to deep learning methods \cite{fan2019interference,Fuchs2020,mun2018deep,Ristea-VTC-2020,rock2019complex}. 

There are many classical RFI mitigation methods \cite{bechter2017dbfmimo, kim2018peer,laghezza2019nxp, mosarim2010d15, uysal2019sync, xu2018orthogonal_noise}, which are usually classified in accordance with the domain in which the interference is mitigated: time, polarization, frequency, code and space. A detailed analysis of these methods is presented in \cite{mosarim2010d15}. When the transmitted signal is a linear chirp waveform, one of the most common method to mitigate interference is to detect, in various ways, the samples of the beat signal affected by interference \cite{laghezza2019nxp} and convert them to zeroes in the time domain. This is commonly known as the zeroing technique. While this approach is fairly simple, it removes part of the useful signal and can become ineffective when the interference has a long duration. 

A series of recent methods~\cite{fan2019interference,Fuchs2020,Ristea-VTC-2020,rock2019complex} rely on deep learning models to mitigate RFI. Rock et al.~\cite{rock2019complex} proposed a convolutional neural network (CNN) to address RFI, aiming to reduce the noise floor while preserving the signal components of detected targets. The authors reported promising results, but they still had concerns regarding the generalization capacity on real data.
Another approach that relies on CNNs is proposed in \cite{Fuchs2020}. The method is based on the U-Net architecture \cite{ronneberger2015u}, performing interference mitigation as a denoising task directly on the range-Doppler map. Fuchs et al.~\cite{Fuchs2020} surpassed classical approaches, but their method fails to fully preserve the phase information. Similarly, Fan et al.~\cite{fan2019interference} proposed a method based on CNNs, their contribution being that of adding residual connections, inspired by the ResNet model \cite{he2016deep}, into the architecture. In another recent work, Ristea et al.~\cite{Ristea-VTC-2020} proposed fully convolutional networks (FCNs) trained on synthetic data samples. The FCNs, as well as other deep models, can estimate the absolute value of range profiles, but are not able to obtain the phase information.

Different from the related works presented above, we employ a decomposition algorithm based on unfolded robust Principal Component Analysis (RPCA) \cite{solomon2019deep}, introducing residual overcomplete auto-encoder blocks in order to obtain an efficient RFI approach. Our model can decompose the radio signal acquired from sensors as a sum of a low-rank matrix (interference signal) and a sparse signal (clear signal with targets). Convergence is achieved by solving a convex minimization problem to retrieve the clear signal, which leads to an iterative principal component pursuit \cite{candes2011robust}. Moreover, as we combine iterative algorithms, which provide a natural recurrent architecture, with residual overcomplete auto-encoders based on convolutional layers, we exploit the benefits of both recurrent and convolutional architectures in order to attain better results. In the experiments, we show that our approach surpasses both FCN models of Ristea et al.~\cite{Ristea-VTC-2020}. Unlike other deep models, which only predict the amplitude, our approach is also able to estimate the phase.

\section{Method}
\label{sec:method}

\subsection{Problem Formulation}

For a continuous-time signal $x(t)$, let $x[n]$ denote the discrete-time signal computed by sampling $x(t)$, $x[n] = x(n \cdot T_S)$, where $T_S$ is the sampling period.

In FMCW radar, the antenna transmits a signal $s_{TX}(t)$, which is a chirp sequence, whose frequency usually follows a saw-tooth pattern. The receive antenna collects $s_{RX}(t)$, which, in the presence of interference from another vehicle, is a mixture of two signals: the signal reflected by targets (having the chirp modulation rate identically with the transmitted signal) and the interference signal (having the chirp modulation rate different from the transmitted signal). Consequently, the received signal is defined as:
\begin{equation}
\label{eq_int_sign}
s_{RX}(t) = \displaystyle\sum_{p=0}^{N_{t}} {A}_{p} \cdot s_{TX}(t-t_{d,p}) + s_{chirp, RFI}(t),
\end{equation}
where ${A}_{p}$ and $t_{d,p}$ are the complex amplitude and the propagation delay of target $p$, respectively, $N_{t}$ is the number of targets, and $s_{chirp,RFI}(t)$ is the interfering signal.

The received signal, $s_{RX}(t)$, is mixed with the  transmitted signal, $s_{TX}(t)$, low-pass filtered and sampled, resulting in the beat signal $s_{b}[n]$. Therefore, $s_{b}[n]$ consists in a sum of complex sinusoids (representing the targets) and an uncorrelated interference $s_{b,RFI}[n]$, which is a chirp signal that is obtained by mixing two chirp signals with different modulation rates. Hence, the beat signal in the presence of uncorrelated interference is written as:
\begin{equation}
\label{eq_beat_sign}
s_{b}[n] = \displaystyle\sum_{p=0}^{N_{t}} {A}_{p} \cdot \exp(2 \pi \cdot j \cdot f_p \cdot n) + 
s_{b,RFI}[n],
\end{equation}
where $j^2=-1$, $f_p = (\alpha \cdot t_{d,p}) \cdot T_S$ is the beat frequency of target $p$ and $\alpha$ denotes the slope of the transmitted chirp. The uncorrelated interference appears as a highly non-stationary component on the beat signal's spectrogram, being spread across multiple frequency bins, as opposed to the signal received from targets, which is present only at some frequency values $f_p$  \cite{alland2019spm}. Hence, we can consider the signals received from targets as narrow band components and the interference as a wide band signal.

We propose to mitigate the interference in the Fourier domain. Therefore, we first apply the Fast Fourier Transform (FFT) to the signal defined in Eq.~\eqref{eq_beat_sign}, obtaining the beat signal spectrum. We consider the FFT of the signal received from targets (the sum in Eq.~\eqref{eq_beat_sign}) as a sparse signal, because there is a limited number of targets, which translates to a few amplitudes on the range profile. The interference signal, $s_{b,RFI}[n]$, is considered a type of noise, because its spectrum contains multiple frequency bins with higher amplitudes. In order to obtain a matrix representation, which is commonly used in deep learning methods, the FFT of each discrete signal is represented as a matrix of shape $N_{FFT} \times 2$, where $N_{FFT}$ is the number of FFT points, and $2$ comes from the real and imaginary parts of FFT, which are viewed as independent vectors. Consequently, we propose a data model composed of a matrix ${L}$ (the FFT of the interference signal) plus a sparse matrix ${S}$ (the FFT of the signal received from targets). Our data model is described in a matrix formulation as:
\begin{equation}
\label{eq_formulation}
{D} = {L}+{S},
\end{equation}
where ${D}$ is the received data, ${L}$ is the interference data and ${S}$ is the clean data with targets. The matrices D, L and S have the same dimension, namely $N_{FFT} \times 2$.

\subsection{Unfolded Robust PCA}

Unfolding \cite{Monga-SPM-2020}, or unrolling an iterative algorithm, was first suggested by Gregor et al.~\cite{gregor2010learning} to accelerate convergence. They showed that by considering each iteration of an iterative algorithm as a layer in a deep network and by concatenating a few such layers, it is possible to train unfolded networks to achieve a dramatic improvement in convergence, significantly reducing the number of training iterations. In the context of RPCA, a principled way to construct learnable pursuit architectures for structured sparse and robust low-rank models was introduced in \cite{sprechmann2015learning}. The proposed networks, derived from the iteration of proximal descent algorithms, were shown to faithfully approximate the solution of RPCA, but the approach was based on a non-convex formulation in which the rank of $L$ was assumed to be known a-priori. This poses a network design limitation, as the rank can vary between different applications or even different realizations of the same application, \ie the number of targets from two signals acquired from the same radio sensor may be different. In contrast, we employ the approach proposed in~\cite{solomon2019deep}, which does not require a-priori knowledge of the rank. 

Unfolding an algorithm can be envisioned as a recurrent neural network, in which the $k^{th}$ iteration is regarded as the $k^{th}$ layer in a feed-forward network. Following~\cite{solomon2019deep}, the $L$ and $S$ matrices for a certain step $k$ are computed as follows:
\begin{equation}
\label{eq_matrix_mul}
\begin{split}
{L}^{k+1} &= SVT_{\lambda_{1}^{k}}\{g_{5}^{k}({L}^{k}) + g_{3}^{k}({S}^{k}) + g_{1}^{k}({D}) \} ,\\
{S}^{k+1} &= \tau_{\lambda_{2}^{k}}\{g_{6}^{k}( {L}^{k}) + g_{4}^{k}({S}^{k}) + g_{2}^{k}({D})\} ,
\end{split}
\end{equation}
where the operator $SVT$ refers to singular value thresholding and the operator $\tau$ and the regularization parameters $\lambda_1, \lambda_2$ are described in \cite{solomon2019deep}. Each function $g_i, \forall i \in \{1,2,...,6\}$ is a transformation, which, in \cite{solomon2019deep}, takes the form of a convolution with a learnable kernel, and, in our approach, takes the form of a more complex function based on residual auto-encoder blocks. The parameters of each $g_i$ are learned independently for each layer $k$. Although, in theory, $L$ is supposed to be a low-rank matrix, we empirically observed that for most data samples, its rank is maximum, \ie equal to $2$. Nevertheless, our empirical results show that unfolded RPCA works well, even if the theoretical constraint regarding the rank of $L$ is not met. For more details about unfolded RPCA, the reader is referred to~\cite{solomon2019deep}. 

\begin{figure}[!t]
\centering
\includegraphics[width=1.0\columnwidth]{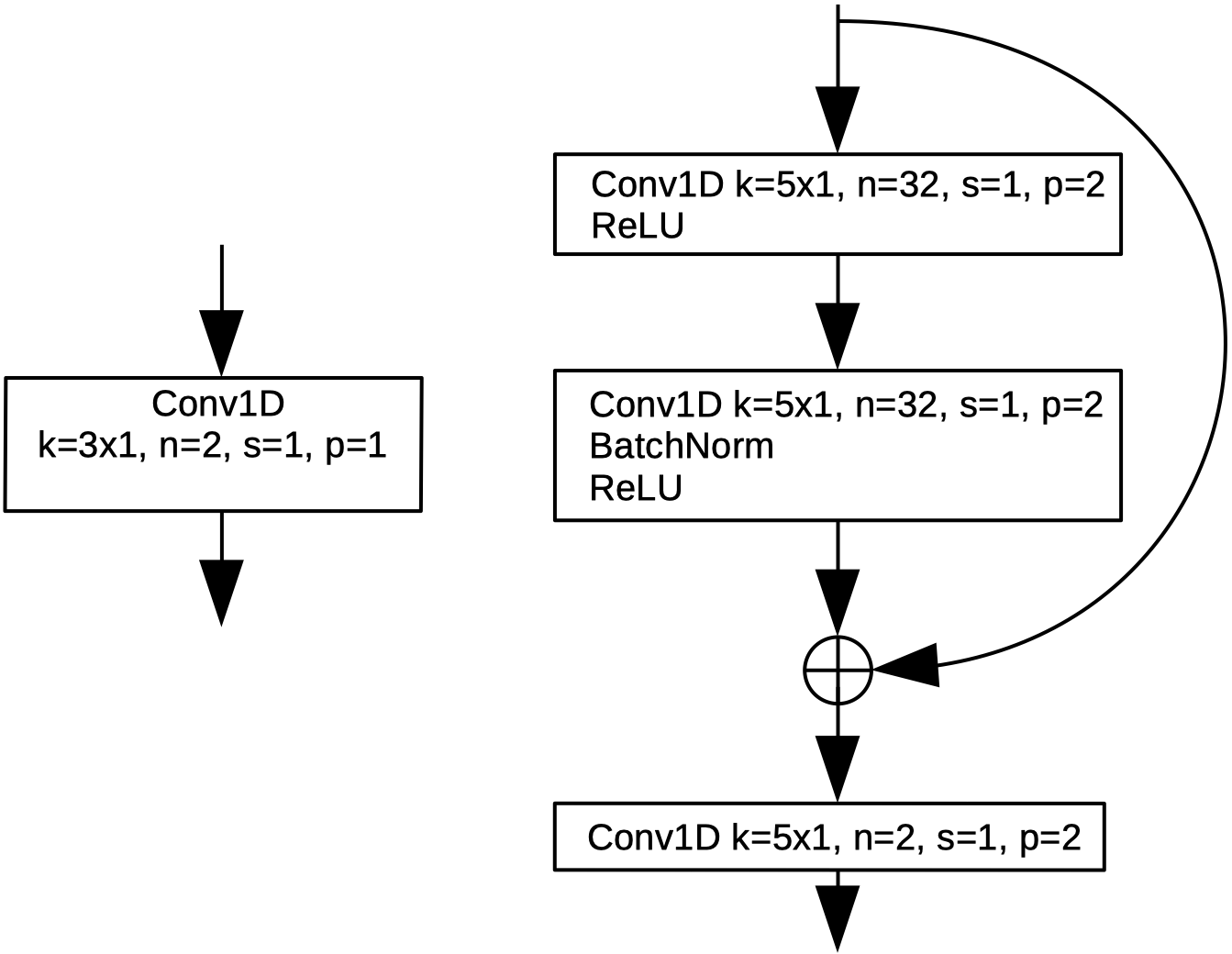}
\caption{Convolutional layer (left) used in the unfolded RPCA of \cite{solomon2019deep} versus our residual overcomplete auto-encoder block (right). The parameters are defined as follows: $k$ is the kernel size, $n$ is the number of filters, $s$ is the stride and $p$ is the padding.}
\label{blocks_fig}
\end{figure}

\subsection{Residual Overcomplete Auto-Encoder Blocks}
In the deep formulation of unfolded RPCA, the recurrent network is based on convolutional layers. Following recent works~\cite{he2016deep,Szegedy-CVPR-2015} showing that deeper models provide better results, we propose to replace the convolutional layers in the deep unfolded RPCA with \textbf{r}esidual \textbf{o}ver\textbf{c}omplete \textbf{a}uto-\textbf{e}ncoder (ROC-AE) blocks. Our block is formed of two convolutional layers with $32$ filters each, and a third convolutional layer with two filters. In a set of preliminary experiments, we observed that some weights converged to $NaN$ values, a problem that is caused by vanishing or exploding gradients. In order to avoid this issue, we insert a skip connection from the input to the third convolutional layer. Our novel block is illustrated in Fig.~\ref{blocks_fig}, in comparison with the approach proposed in \cite{solomon2019deep}, which is based on a single layer of convolution. Both architectures use tensors of $1\times2048\times2$ components as input and output, respectively. We note that the first two convolutional layers in our block have $32$ filters, generating activation maps of $1\times2048\times32$ components. Hence, our residual blocks are designed like overcomplete auto-encoders (the latent space is higher than the input space). In the experiments, we show that our residual overcomplete auto-encoder blocks significantly outperform \textbf{r}esidual \textbf{u}nder\textbf{c}omplete \textbf{a}uto-\textbf{e}ncoder (RUC-AE) blocks with equivalent depth.


\section{Experiments}
\label{sec:experiments}

\subsection{Data Set}
The automotive radar inference mitigation (ARIM) data set \cite{Ristea-VTC-2020} is a large scale database consisting of 48,000 radio signal samples, synthetically generated while trying to replicate realistic automotive scenarios with one source of interference. The data set is split into a training set of 40,000 samples and a test set of 8,000 samples. We split the training data into two disjunctive sets for training (32,000 samples) and validation (8,000 samples), according to Ristea et al.~\cite{Ristea-VTC-2020}.

Each sample is generated using randomly selected values for the following parameters: signal-to-noise ratio (SNR), signal-to-interference ratio (SIR), relative interference slope, number of targets, amplitude, phase and distance of each target. 
One of the biggest advantages that are provided by the ARIM data set is that we have access to clean and perturbed signal pairs. Therefore, we are able to use the model described in Section~\ref{sec:method}, as it requires access to both interference and clean target signals during training.

To our knowledge, ARIM \cite{Ristea-VTC-2020} is the only large scale data set that is publicly available for the radar interference mitigation task. Therefore, we evaluate our deep learning method against other competing approaches only on ARIM.

\begin{table*}[t]
\renewcommand{\arraystretch}{1.15} 
\centering
\noindent
\caption{Validation and test results on the ARIM data set attained by various versions of unfolded RPCA versus an oracle based on true labels, zeroing and state-of-the-art FCN models~\cite{Ristea-VTC-2020}, respectively. The best results (excluding the oracle) are highlighted in bold. The symbol $\uparrow$ means higher values are better and $\downarrow$ means lower values are better.} 
\vspace{0.1cm}
\setlength\tabcolsep{2.0pt}
\begin{tabular}{|l|cccc|cccc|ccc|}
\hline
 & \multicolumn{4}{|c|}{{Validation set}} & \multicolumn{4}{|c|}{{Test set}} & \multicolumn{3}{|c|}{{Inference time}}\\ 
\cline{2-12}
{{Method}} & $\overline{\Delta \mbox{SNR}}\uparrow$ & {AUC}$\uparrow$ & {MAE}$\downarrow$& {MAE}$\downarrow$ & $\overline{\Delta \mbox{SNR}}\uparrow$ & {AUC}$\uparrow$ & {MAE}$\downarrow$& {MAE}$ \downarrow$ & CPU & GPU & NX\\
 &  & & {(dB)} & {(degrees)} & & & {(dB)} & {(degrees)} & (ms)& (ms) & (ms)\\ 
\hline
\hline
Oracle & 12.92 & 0.978 & 0 & 0 & 13.08 & 0.978 & 0 & 0 & - & - & - \\ 
\hline
{Zeroing} & 5.27 & 0.951 & 1.26 & 6.80 & 5.44 & 0.951 & 1.27 & 6.79 & $<$1& $<$1 & -\\ 
{Shallow FCN~\cite{Ristea-VTC-2020}} & 10.34 & 0.965 & 2.20 & - & 10.49 & 0.965 & 2.21 & - & 471 & 62 & - \\ 
{Deep FCN~\cite{Ristea-VTC-2020}} & \textbf{12.90} & 0.972 & 1.21 & - & \textbf{13.06} & 0.972 & 1.22 & - & 8400 & 66 & - \\ 
{Unfolded RPCA \cite{solomon2019deep}} & 12.14 & 0.968&1.47 &5.58 &12.33 &0.970 &1.47 &5.56 & 55 & 45 & 273\\ \hline
{Unfolded RPCA RUC-AE} & 9.58 & 0.967 & 2.83 & 5.04 & 9.83 & 0.966 & 2.76 & 5.71 & 76 & 35 & 242\\ 
{Unfolded RPCA ROC-AE} (ours) & 10.15 &\textbf{0.975} & \textbf{0.53} & \textbf{2.45} & 10.46 & \textbf{0.976} & \textbf{0.55} & \textbf{2.55} & 122 & 40 & 299\\ 
\hline
\end{tabular}
\label{tab_results_arim}
\end{table*}

\subsection{Evaluation Metrics}
Typically, the goal in radar signal processing is to maximize the target detection performance. Therefore, an intuitive metric is the area under the receiver operating characteristics curve (AUC), which describes the ability to disentangle targets from noise at various thresholds. 
Another performance indicator is the mean absolute error (MAE) in decibels (dB) between the range profile amplitude of targets computed from label signals and the amplitude of targets from predicted signals. 
In radar signal processing, recovering the phase of targets is equally important, because it is essential in estimating other mandatory parameters, \eg target velocity. Thus, we also report the MAE in degrees between the range profile phase of targets computed from label signals and the phase of targets from predicted signals. 
In our evaluation, we additionally report 
the mean SNR improvement ($\overline{\Delta \mbox{SNR}}$), which is computed for the target with the highest amplitude in a signal, as the difference between SNR before and after interference mitigation.

\begin{figure}[!t]
\centering
\includegraphics[width=1.0\columnwidth]{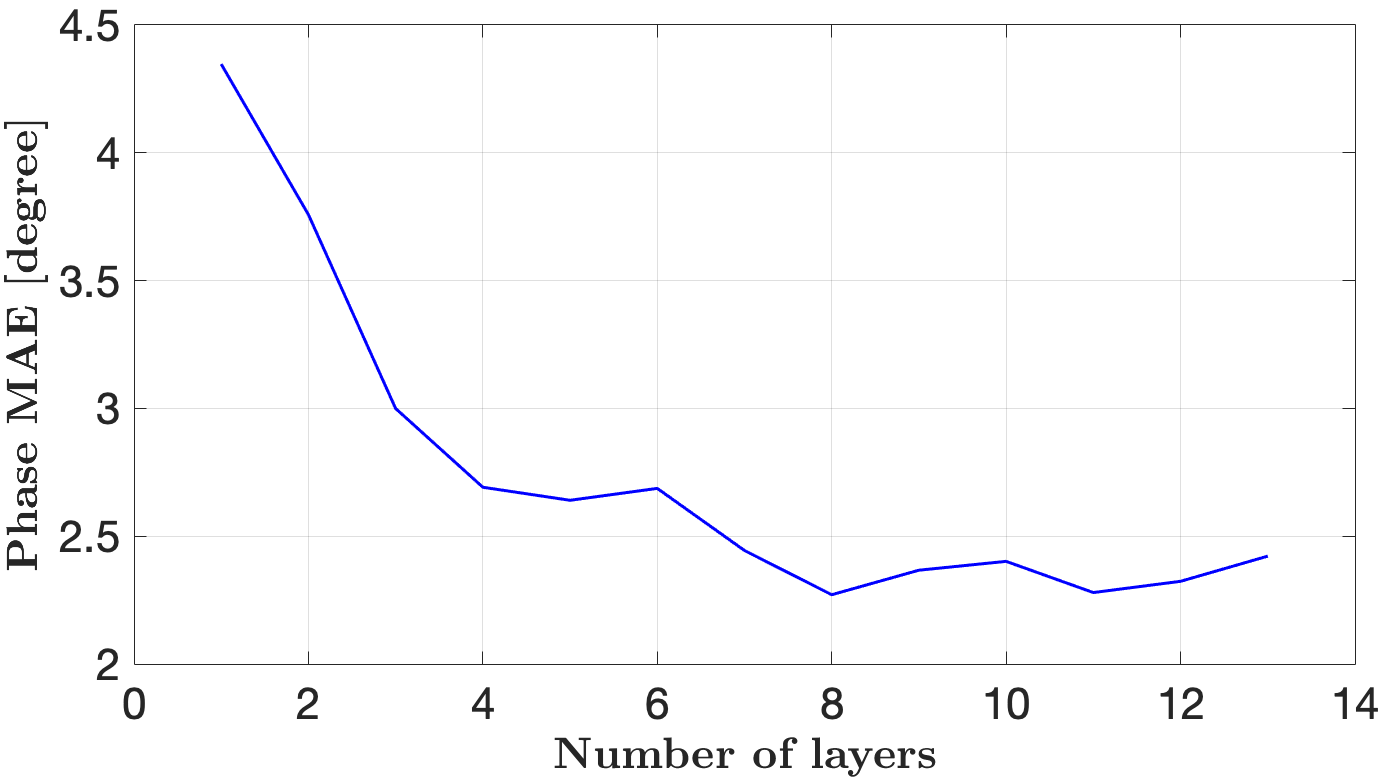}
\caption{The influence of the number of neural network layers (from 1 to 13) on the phase MAE measured in degrees.}
\label{fig_1}
\end{figure}

\subsection{Hyperparameter Tuning}
In order to minimize the chance of overfitting in hyperparameter space, we tune our hyperparameters on the validation set. The number \textit{k} of unfolded network layers was validated on the evaluation set, considering values from 1 to 13. As shown in Fig. ~\ref{fig_1}, the optimal value is $k=8$. Regarding the training process, we trained our network for $100$ epochs with mini-batches of $20$ samples using the Adam optimizer \cite{Kingma-ICLR-2015} with a learning rate of $5 \cdot 10^{-4}$ and a weight decay of $10^{-6}$. We also added a learning rate scheduler with a step of $30$ epochs and a decay factor of $0.5$.

\subsection{Quantitative Results}

We compare the unfolded RPCA models based on ROC-AE or RUC-AE blocks with the most common classical approach, called zeroing, two FCNs described in \cite{Ristea-VTC-2020}, the unfolded RPCA approach proposed in \cite{solomon2019deep} and an oracle computed from the ground-truth labels. We present the corresponding results in Table~\ref{tab_results_arim}. 
Our unfolded RPCA model based on ROC-AE blocks outperforms the zeroing method by a large margin, in terms of all performance measures. When comparing the model based on standard (undecomplete) AE blocks with the one based on overcomplete AE blocks, we observe that the latter model attains superior results, regardless of the metric. The unfolded RPCA based on undercomplete AE blocks attains lower scores even compared with the unfolded RPCA \cite{solomon2019deep}. This clearly shows the necessity to use overcomplete AE blocks in order to obtain performance improvements. In terms of AUC and amplitude MAE, the unfolded RPCA \cite{solomon2019deep} is below the FCN models \cite{Ristea-VTC-2020}. The introduction of the ROC-AE blocks brings significant performance gains to the unfolded RPCA model, surpassing all models in terms of AUC, amplitude MAE and phase MAE. 
Even if our approach attains inferior results in terms of $\overline{\Delta \mbox{SNR}}$ compared with both FCN models \cite{Ristea-VTC-2020}, the latter models cannot estimate the phase of targets. This is a major drawback of the FCN models. We emphasize that neither version of unfolded RPCA suffers from this problem.

Additionally, we observe that our model obtains lower performance in terms of the mean SNR improvement, because of the data samples having a small signal-to-interference ratio (SIR). A small SIR implies that the FFT of the signal does not meet the sparsity condition, because the targets are close to the noise floor. Having targets near or in the noise floor at training time may interfere with the proposed data modeling.

\begin{figure}[!t]
\centering
\includegraphics[width=1.0\columnwidth]{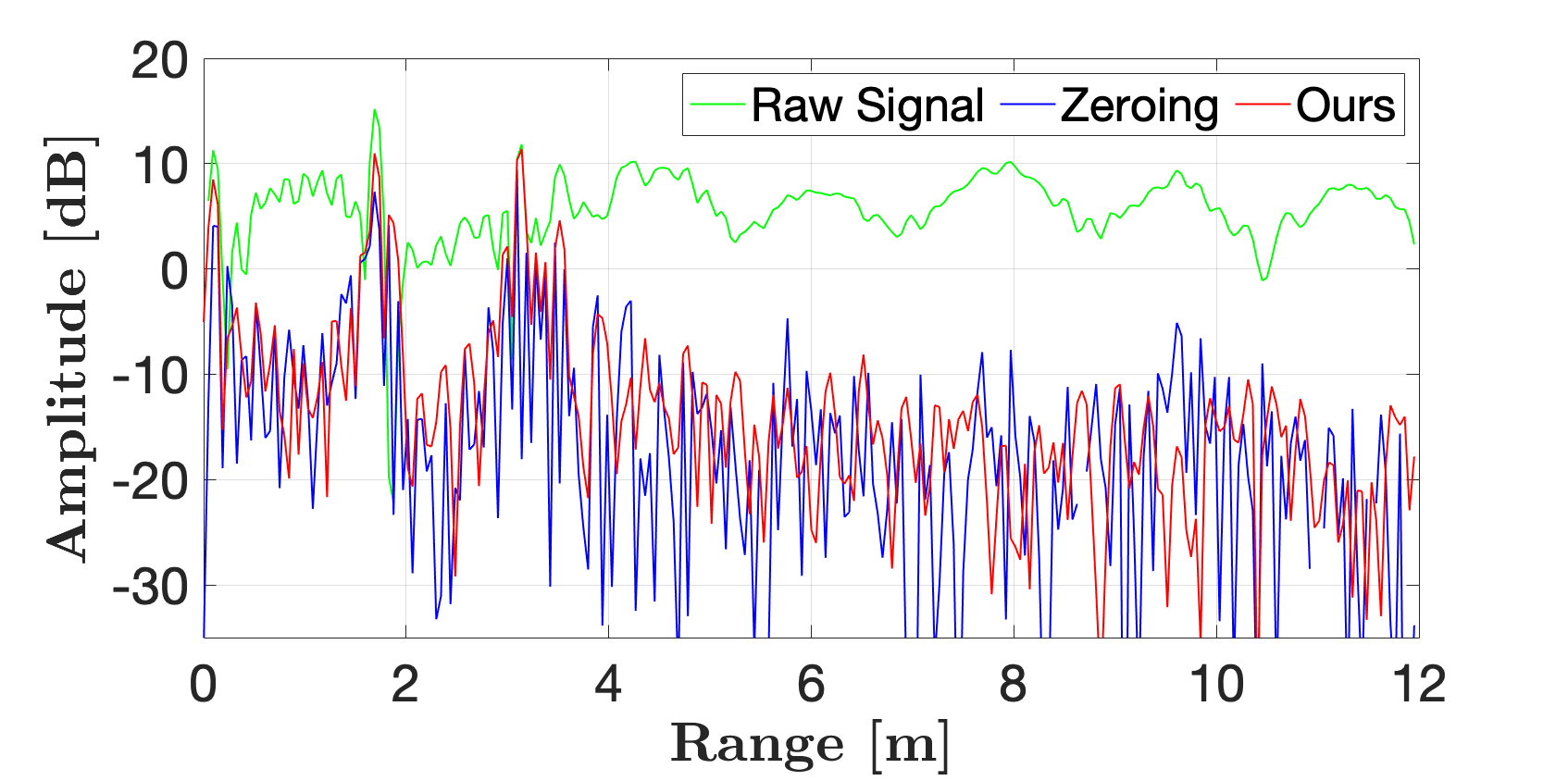}
\caption{The beat signal spectrum for RFI mitigation results with our unfolded RPCA based on ROC-AE blocks versus the zeroing method on a real beat signal spectrum. For reference, we also added the input signal captured by the NXP TEF810X 77 GHz radar transceiver.}
\label{fig_2}
\end{figure}

\subsection{Running Time}

In the radar signal domain, a key element is the capability of an algorithm to process data in real-time on low-power processing units, \eg embedded devices. Therefore, we analyze the inference time of each method, \ie the average time required to process a time domain signal and output the corresponding beat signal spectrum. 

The CPU and GPU times reported in Table~\ref{tab_results_arim} were measured on a machine with Intel Core i7 CPU and NVIDIA RTX 2080Ti GPU. As expected, the zeroing method has the best inference time, but its accuracy levels are much lower compared to our deep learning approach. Moreover, we observe that our unfolded RPCA based on ROC-AE blocks is quicker than both FCNs, especially on CPU, while also offering better results. The algorithm proposed in \cite{solomon2019deep} is slightly faster than our approach because of its shallower architecture.

In addition, we measured the running times of the three unfolded RPCA models on an Nvidia Jetson Xavier NX embedded system. The corresponding time measurements are reported in the last column of Table~\ref{tab_results_arim}. We observe that our unfolded RPCA based on ROC-AE blocks requires about $22$ extra milliseconds on the lower-end GPU with respect to the baseline unfolded RPCA. We conclude that our unfolded RPCA based on ROC-AE blocks provides the optimal trade-off between accuracy and speed. With the reported times, our approach is a viable candidate to be deployed on embedded devices placed on board road vehicles.

\subsection{Qualitative Results}

In addition to the results on ARIM, we assessed the generalization capacity of our approach on real data, by testing it on samples provided by the NXP company, which were captured with the NXP TEF810X 77 GHz radar transceiver, in real-world scenarios. In Fig.~\ref{fig_2}, we show an example of interference mitigation performed by our unfolded RPCA based on ROC-AE blocks on a real radar signal in comparison with the zeroing algorithm. We underline that the shown example contains three close-range targets, at roughly 0, 2 and 3 meters. Both models seem capable of reducing the noise floor, but our approach is better at estimating the targets. More precisely, the amplitude around the targets is higher for our approach compared to zeroing.

\section{Conclusion}
\label{sec:conclusions}

In this paper, we introduced an unfolded robust PCA model based on residual overcomplete auto-encoder blocks for automotive radar interference mitigation, which is capable of estimating both the magnitude and the phase of automotive radar signals. We compared our model with several baseline approaches in a comprehensive experiment, showing that our model provides superior results in terms of accuracy and time. We also showed the real-time processing capability of our approach, as well as its generalization capacity on real data. In future work, we aim to analyze the scenario with multiple interference sources.

\section*{Acknowledgments}

The authors thank reviewers for their useful feedback leading to improvements of this work. The authors would also like to thank Adrian \c{S}andru from SecurifAI for helping to measure the inference times of the unfolded RPCA models on Jetson NX. The authors acknowledge national funding authorities and the ECSEL Joint Undertaking, which funded the PRYSTINE project under grant 783190. This work was co-funded through the Competitiveness Operational Program 2014-2020, Axis 1, contract no. 3/1.1.3H/24.04.2019, MySMIS ID: 121784. This article has also benefited from the support of the Romanian Young Academy, which is funded by Stiftung Mercator and the Alexander von Humboldt Foundation for the period 2020-2022.

{\small
\bibliographystyle{ieee_fullname}
\bibliography{refs}
}

\end{document}